# Phishing – A Growing Threat to E-Commerce

*M. Tariq Banday\** and *Jameel A. Qadri\*\**

*\* Department of* Electronics & Instrumentation Technology,
The University of Kashmir, Srinagar – 190006,
email: sgrmtb@yahoo.com.
\*\* School of Computing, Middlesex University,
Hendon, London, UK,
email: scorpiojameel@yahoo.com.

**Abstract**: *In today's business environment, it is difficult to imagine a workplace without access to the web, yet a variety of email born viruses, spyware, adware, Trojan horses, phishing attacks, directory harvest attacks, DoS attacks, and other threats combine to attack businesses and customers. This paper is an attempt to review phishing – a constantly growing and evolving threat to Internet based commercial transactions. Various phishing approaches that include vishing, spear phishng, pharming, keyloggers, malware, web Trojans, and others will be discussed. This paper also highlights the latest phishing analysis made by Anti-Phishing Working Group (APWG) and Korean Internet Security Center.*

## Introduction

Electronic Commerce (E-Commerce) is commercial transactions conducted electronically especially using a computer over a large network like Internet. It involves exchange of business information using electronic data interchange (EDI), email, electronic bulletin boards, fax transmissions, electronic funds transfer, etc. Internet shopping, online stock and bond transactions, selling and purchase of soft merchandise like documents, graphics, music, software, and other customer/business to business/customer transactions are one or the other forms of e-commerce (Michael-2002). E-commerce allows customers a prescribed admission to a host computer and lets them serve themselves. E-commerce has given a boon to both customers and businesses by driving down costs and prices. E-commerce allows real-time business across geographical borders round the clock. In developed countries almost all business employs e-commerce or has e-commerce provisions and in developing countries like India, it is registering a rapid growth in terms of both popularity among consumers and the revenue generated through e-commerce (Vashitha–2005). This virtual market is facing a continuously growing threat in terms of identity theft that is causing short-term losses and long-term economic damage. Among several identity thefts, phishing and its various variants are the most common and deterrent to e-commerce. The scams and frauds on Internet like identity theft leads to the problem of computer theft and massive computer penetration and espionage. Phishing is an outcome of unsolicited bulk email (UBE) and unsolicited commercial email (UCE) also referred to as spam. Unfortunately some companies realized that not only could they communicate by email with staff and existing business partners but they could also reach out to millions of potential new customers on the web, introducing themselves and their services for minimum cost and required only a tiny response and service uptake to make it all worthwhile. Email spam has become a fact of life. A variety of new message threats now combine to attack individuals, organizations, and businesses and thus prove to be a grave threat to e-commerce, they include email born viruses, spyware, adware, Trojan horses, directory harvesting attacks (DHA), denial of services (DoS) attacks and more importantly phishing attacks.

## Phishing

Phishing is a term used to describe spoof emails and other technical ploys to trick receipts into giving up their personal or their company's confidential information such as social security and financial account credentials and other identity and security information. This form of identity theft employs both social engineering and technical subterfuge to steal account access information, identity or other proprietary information that can be sold on to third party via specialized chat rooms established specifically for the purpose of selling such information. Selling the information on reduces the risk of being apprehended by minimizing the direct link between the hacker and those using the information to gain unauthorized access to accounts and profiting from them. Social engineering schemes use spoofed emails to lead customers to counterfeit websites designed to trick recipients into divulging financial data. Technical subterfuge schemes plant crime ware onto computers to steal credentials directly using keylogging systems. Pharming crimeware misdirects users to fraudulent sites or proxy servers, typically through Domain Name System (DNS) hijacking or poisoning. The term "phishing" evolved from the word "fishing" and follows a very similar approach. Fraudsters and scammers, the "fishermen", send out large quantities of deceptive emails, the "bait", to mostly random address across the Internet. Some phishing emails look extremely professional and realistic, while others are crude and badly constructed but have a common goal - to steal information through deception.

Phishing is a serious and increasingly prolific form of spam, and is one of the main tactics employed in business and consumer identity theft. Phishing actually comprises of two online identity thefts used together. In phishing scams, the identity of the target company - commonly a bank, online payment service, or other reputable business - is stolen first in order to steal even more identities: those of unsuspecting customers of the targeted company. A typical phishing attack is made up of two components: an authentic-looking email and a fraudulent Web page. This form of spam frequently uses professional-looking, HTML-based emails that include company logos, colors, graphics, font styles, and other elements to successfully spoof the supposed sender. The content of the phishing email is usually designed to confuse, upset, or excite the recipient. Typical email topics include account problems, account verifications, security updates/upgrades, and new product or service offerings. Recipients of the email are prompted to react immediately. They then click on a link provided in the email body, which actually directs them to the phishing Web page. The intent is to lure recipients into revealing sensitive information such as usernames, passwords, account IDs, ATM PINs, or credit card details. Like the phishing email, the phishing Web page almost always possesses the look and feel of the legitimate site that it copies, often containing the same company logos, graphics, writing style, fonts, layout, and other site elements. This spoofed Web page may also include a graphical user interface (GUI) intended to lure the user into entering their bank account information, credit card number, social security number, passwords, or other sensitive information. Either the phisher, or an anonymous remote user that is sent the information, can then use the stolen information.

The phishing "ecosystem" consists of a collection of individuals who play various roles within the phishing space, ranging from the financially-motivated botnet creators to those who actively pursue and prosecute the cybercriminals. In this ecosystem, a large industry of buying and selling - a "microeconomy" - exists within the phishing underground, involving botnet (Biever-2004) creators, perpetrators, and enablers. However, these three player groups are complex and intertwined: a single individual or multiple perpetrators can play separate or simultaneous roles (Trend-2006).

*Types of Phishing Attacks*

Phishing approaches used for identity thefts are constantly growing and new variants are tried and used to attack business organizations, financial institutions, and customers. Some of the most prevalent types of phishing attacks are presented hereunder.

***Deceptive Phishing*** is the most common broadcast method today and involves sending messages about the need to verify account information, system failure requiring users to re-enter their information, fictitious account charges, undesirable account changes, new free services requiring quick action, and many other scams are broadcast to a wide group of recipients with the hope that the unwary will respond by clicking a link to or signing onto a bogus site where their confidential information can be collected.

*Malware - Based Phishing* refers to scams that involve users to unknowingly running malicious software on their PCs. Malware is as an email attachment, as a downloadable file from a web site, or by exploiting known security vulnerabilities - a particular issue for small and medium businesses (SMBs) who are not always able to keep their software applications up to date.

*Keyloggers and Screenloggers* are particular varieties of malware that track keyboard input in the backdoor and send relevant information to the hacker via the Internet. They embed themselves into web browsers as small utility programs known as helper objects that run automatically when the browser is started as well as into system files as device drivers or screen monitors.

*Session Hijacking* is that phishing attack in which users' activities are monitored until they sign in to a target account or transaction and establish their bona fide credentials. At that point the malicious software takes over and undertakes unauthorized actions, such as transferring funds, without the user's knowledge.

*Web Trojans* are a type of pop up running invisibly on user machines. When users are attempting to log in, they collect the user's credentials locally and transmit them to the phisher.

*Hosts File Poisoning* involves changing the host files of the operating system that contain the IP addresses corresponding to the web addresses. When a user types a URL to visit a website it must first be translated into an IP address before it's transmitted over the Internet. The majority of SMB users' PCs running a Microsoft Windows operating system first look up for these "host names" in their "hosts" file before undertaking a Domain Name System (DNS) lookup. By "poisoning" the hosts file, hackers have a bogus address transmitted, taking the user unwittingly to a fake "look alike" website where their information can be stolen.

*System Reconfiguration Attacks* modify settings on a user's PC for malicious purposes. For example: URLs in a favorites file might be modified to direct users to look alike websites. For example: a bank website URL may be changed from "mybank.com" to "mybanc.com".

*Data Theft* refers to stealing subset of sensitive information stored locally on unsecured PCs which actually is stored elsewhere on secured servers. Certainly PCs are used to access such servers and can be more easily compromised. Data theft is a widely used approach to business espionage. By stealing confidential communications, design documents, legal opinions, and employee related records, etc., thieves profit from selling to those who may want to embarrass or cause economic damage or to competitors.

***DNS-Based Phishing also called Pharming*** is a term given to hosts file modification or Domain Name System (DNS) based phishing. With a pharming scheme, hackers tamper with a company's hosts files or domain name system so that requests for URLs or name service return a bogus address and subsequent communications are directed to a fake site. This results in users falling unwarily victims by working on the websites controlled by hackers where they enter confidential information.

***Content-Injection Phishing*** is used to describe the situation where hackers replace part of the content of a legitimate site with false content designed to mislead or misdirect the user into giving up their confidential information to the hacker. For example, hackers may insert malicious code to log user's credentials or an overlay which can secretly collect information and deliver it to the hacker's phishing server.

***Man-in-the-Middle Phishing*** is harder to detect than many other forms of phishing. In these attacks hackers position themselves between the user and the legitimate website or system. They record the information being entered but continue to pass it on so that users' transactions are not affected. Later they can sell or use the information or credentials collected when the user is not active on the system.

***Search Engine Phishing*** occurs when phishers create websites with attractive offers and have them indexed legitimately with search engines. Users find the sites in the normal course of searching for products or services and are fooled into giving up their information. For example, scammers have set up false banking sites offering lower credit costs or better interest rates than other banks. Victims who use these sites to save or make more from interest charges are encouraged to transfer existing accounts and deceived into giving up their details.

***Spear Phishing*** is a colloquial term that can be used to describe any highly targeted phishing attack. Spear phishers send spurious emails that appear genuine to a specifically identified group of Internet users, such as certain users of a particular product or service, online account holders, employees or members of a particular company, government agency, organization, group, or social networking website. Much like a standard phishing email, the message appears to come from a trusted source, such as an employer or a colleague who would be likely to send an email message to everyone or a select group in the company (e.g., the head of human resources or a computer systems administrator). Because it comes from a known and trusted source, the request for valuable data such as user names or passwords may appear more plausible. Whereas traditional phishing scams are designed to steal information from individuals, some spear phishing scams may also incorporate other techniques, ranging from computer hacking to "pretexting" (the practice of getting personal information under false pretences), to obtain the additional personal information needed to target a particular group or to enhance the phishing emails' credibility. In essence, some criminals will use any information they can to personalize a phishing scam to as specific a group as possible (Microsoft-2005). A recent scam of this nature was encountered at AT&T; a very large telecommunications company (Lazarus-2006).

***Vishing*** or ***Voice phishing*** can work in two different ways. In one version of the scam, the consumer receives an email designed in the same way as a phishing email, usually indicating that there is a problem with the account. Instead of providing a fraudulent link to click on, the email provides a customer service number that the client must call and is then prompted to "log in" using account numbers and passwords. The other version of the scam is to call consumers directly and urge them to call the fraudulent customer service number immediately in order to protect their account. Vishing criminals may also even establish a false sense of security in the consumer by "confirming" personal information that they have on file, such as a full name, address or credit card number (FCAC-2006). Vishing poses a particular problem for two reasons. First, criminals can take advantage of cheap, anonymous Internet calling available by using Voice over Internet Protocol (VoIP), which also allows the criminal to use simple software programs to set up a professional sounding automated customer service line, such as the ones used in most large firms. Second, unlike many phishing attacks, where the legitimate organization would not use email to request personal information from accountholders, vishing actually emulates a typical bank protocol in which banks encourage clients to call and authenticate information (Schulman-2006).

### Growth of Phishing Scams

An estimated one million computers are under the control of hackers worldwide. German security analysts at Aachen University reperted more than 100 botnets in three months, which ranged in size from a few hundred compromised computers to 50,000 machines (Holz–2005). The effects caused by phishing on business and customers are far reaching and include substantial financial loss, brand reputation damage, lost customer data files, possible legal implications, significant decrease in employee productivity, improper IT resource utilization, and other administrators impacts. Besides direct financial loss, erosion of public trust in the Internet is a direct impact of fishing on electronic trading. Over the past several years, law enforcing officials have successfully apprehended, prosecuted (Stevenson, Robert Louis B-2005), and convicted phishers and bot herders. However, apprehending these criminals is becoming increasingly difficult as they become more professional and sophisticated in their operations. Organizations such as the Anti-Phishing Working Group (APWG), the Federal Trade Commission, Digital Phishnet, Korean Internet Security Center, and others have initiated collaborative enforcement programs to combat phishing and identity theft. These industry groups - many of which are global and pan-industrial in scope - focus on numerous areas, including identifying phishing Web sites, sharing best practice information, aiding criminal law enforcement, and assisting in apprehending and prosecuting those responsible for phishing and identity theft crimes. Eight of the top 10 U.S. banks belong to the APWG; its network of global research partners includes some of the world's top e-commerce associations and watchdogs.

According to Korea Phishing Activity Trends Report, published by Korean Internet Security Center (KrCERT/CC-2006), in the year 2006 the total number of phishing sites reported is 1,266, and the average number per month is 105.5. Compared with the average number of phishing sites in 2005, it can be inferred that the recent average number of phishing sites is little bit increased, and that the hacking attempts with the financial objectives are increasing. The second half of year 2006 has many cases so-called "ROCK Phish", in which one IP system hosts many different domain names. This means the number of targeted financial institutions is far more than the number of IP of the phishing hosts. As per the same report most of the phishing attacks used port 80. Further, the most hijacked brands reported are online marketplace company (e-commerce) and online payment gateway company (financial service) in November 2006. These two brands take 46% of all phishing. The report also indicated the most targeted industry sector for phishing attacks is financial service with e-commerce remaining the second (Chat I).

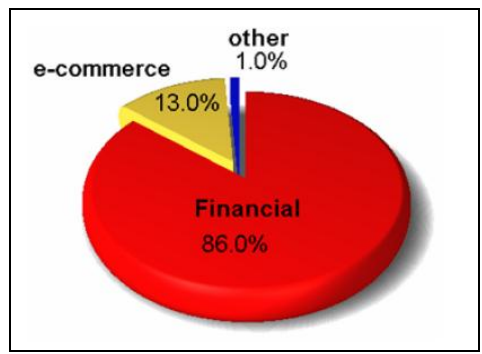

**Chat I: Countries to which Hijacked brands belong**
**Source: Korean Phishing Activity Report**

The financial services sector occupies 86.0% and e-commerce 13% of all hijacked brands in December 2006 and most hijacked brands belong to United States (83%) (Chat II).

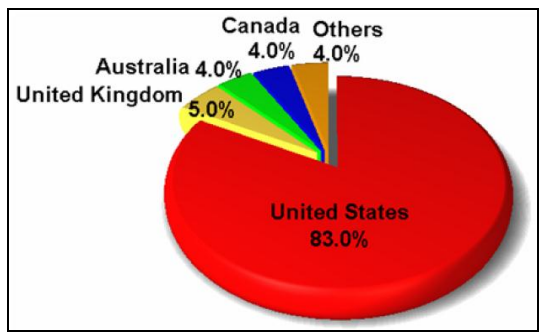

**Chat II: Hijacked Brands by Industry in Dec, 2006.**

Source: Korean Phishing Activity Report

The recent main stream of cyber attacks is for the financial gain and spam emails are also increasing. In addition, many insecure servers are hacked and exploited as phishing hosts for deceiving users to input their personal and financial information. Therefore, every Internet users should use more caution for the possibility of financial loss by Internet frauds. As per the Anti-Phishing Working Group (APWG) report phishing attacks have increased significantly and have reached the highest number of 29930 in January, 2007 (Chat III). This count is the phishing email reports received by the APWG from the public and its member organizations and its research partners.

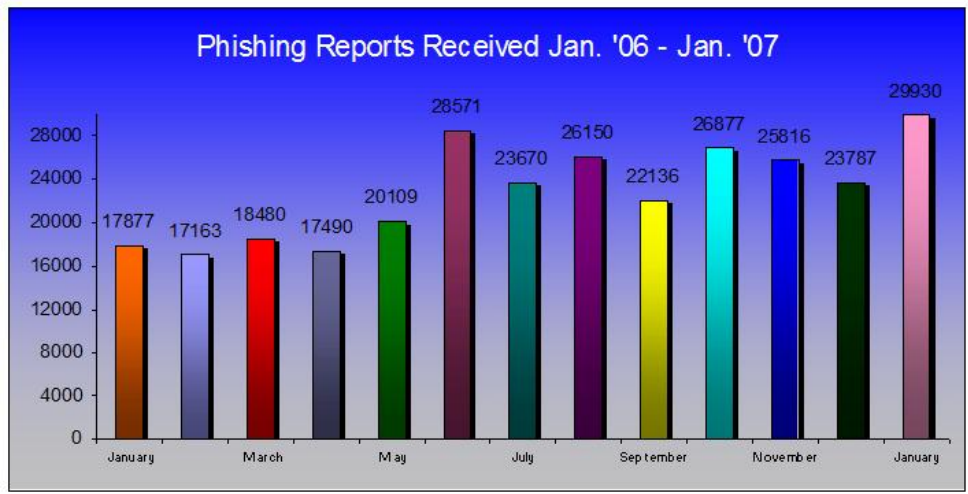

**Chat III: Phishing Reports Received by APWG from January, 2006 to January 2007**
*Source: Anti-Phishing Working Group (APWG)*

The count of phishing reports as reported by APWG is very alarming owing to the fact that companies that are victimized by phishing may not report these instances to law enforcement. Unlike some other types of Internet-based crime, such as hacking, that may be conducted surreptitiously, phishing, by its nature, involves public misuse of legitimate companies' and agencies' names and logos. Nonetheless, some companies may be reluctant to report all such instances of phishing to law enforcement - in part because they are concerned that if the true volume of such phishing attacks were made known to the public, their customers or accountholders would mistrust the companies or they would be placed at a competitive disadvantage.

Though the number of unique phishing websites detected by APWG has decreased in December, 2006 and January 2007, yet the total average number of such websites in year 2006 has increased significantly in comparison to that in the year 2005. Chat IV illustrates the growth of new phishing websites from January 2006 to January 2007.

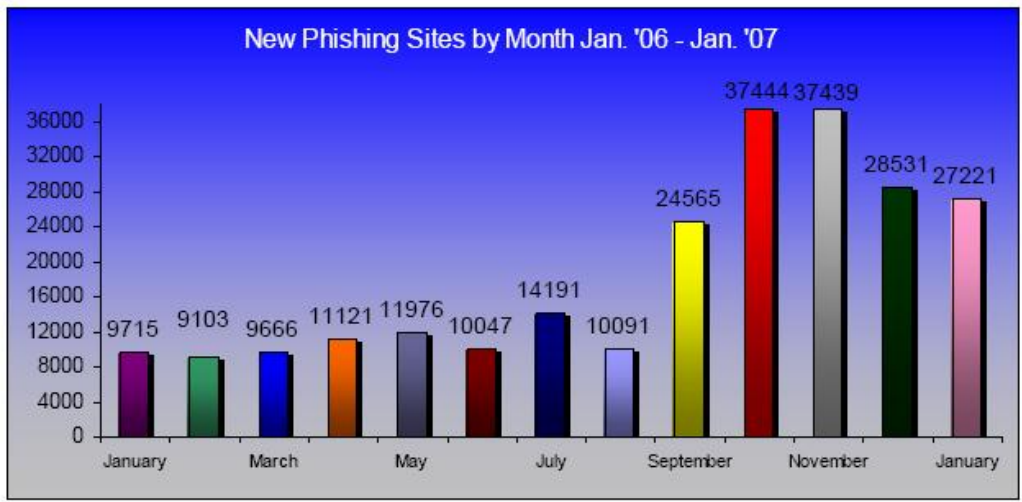

**Chat IV: New Phishing websites detected by APWG from January, 2006 to January 2007**
**Source: Anti-Phishing Working Group (APWG)**

Most IT decision makers at Small and Medium Businesses are aware of phishing through consumer scams but many do not believe that phishing poses a threat to organizations that are neither in financial services sector nor in the public eye. Phishing technologies have become so advanced that no one is immune to the possibilities of being scammed if they are searching for goods and services on the web and intend to pay by credit card (Dan Ferguson, 2006).

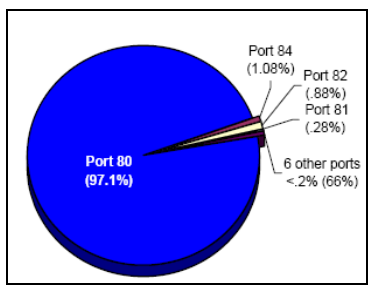

**Chat V: Ports used by Phishers**
**Source: Anti-Phishing Working Group (APWG)**

Software processes running in computers need software ports to connect with other software processes. Inter Process Communication (IPC) takes place between these ports. Each process must be distinguished from other process and this is done with port numbers. Ports are numbered form 0 to 65536 with the most popular application and protocols such as FTP, SMTP, SNMP, HTTP, etc. are pre-assigned to "Well Known" ports. When a web browser contacts a web server it directs its message to port 80, "Well Known Port for HTTP services". APWG report for January, 2007 indicated that a continuing trend of HTTP port 80 being the most used port for all phishing sites that have been reported (Chat V).

Chat VI shows the number of brands hijacked, number of unique password stealing malicious code applications and number of password stealing malicious URLs from January 2006 to January 2007. Though hijacked brands remained steady throughout these months yet numerous non-traditional websites such as social networking portals and gambling sites were hijacked.

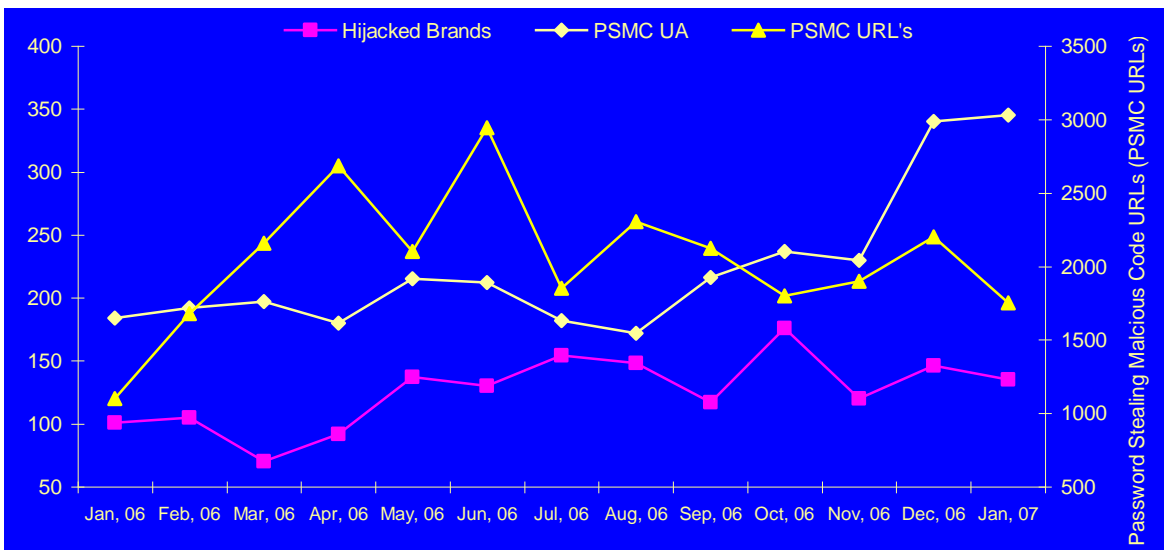

**Chat VI: Hijacked Brands, Password Stealing Malicious Code Applications and URLs**
**Source Data: Anti-Phishing Working Group (APWG)**

Chat VI also shows the number of phishing based Trojans as reported by APWG in form of malicious code applications and URLs that monitor specific actions for identity thefts.

Without paying careful attention to all suspicious emails and carefully checking the validity of all websites requiring entry of confidential information, employees of any size of company may be easily persuaded to give up either personal or company confidential information.

The existence of underground phishing ecosystems and the large financial gains through botnets have transformed phishing into worldwide organized crime. Profits for these cybercriminals - as observed through monitored chat room discussions, apprehensions, trials, and convictions - have ranged from tens of thousands to millions of dollars. Yet, businesses and consumers are greatly impacted by significant financial losses and other short and long-term damage to overall financial health, brand, and reputation. Businesses and consumers can protect themselves from the devastating effects of phishing due to botnet activities in two ways: educating themselves about phishing techniques and employing technology solutions that combat phishing. The Anti-Phishing Working Group like other such groups, list several guidelines that help individuals and organizations to educate themselves and thus avoid falling victim to a phisher.

## Conclusion

Phishing is relatively new, complex and continuously evolving phenomenon that includes social engineering as well as malicious technology. There is no fool proof technology or legislative system that can protect against or prevent phishing attacks from getting through. However, properly deployed combinations of technology coupled with employee education and diligence can significantly reduce the likelihood that a business or a customer will succumb and fall victim to this growing threat. Further, public reporting mechanism established by governments and the law enforcement community with participation from the private sector have proved to greatly understand the scope and magnitude of phishing and identity theft.